\documentclass[aps,prc,twocolumn,superscriptaddress,amsfonts,amsmath,amssymb,showpacs,floatfix]{revtex4-1}

\usepackage{graphicx}
\usepackage{longtable}
\usepackage{bm}
\usepackage{multirow}
\usepackage{hyperref}
\usepackage{ulem}
\hypersetup{colorlinks,citecolor=blue,filecolor=blue,linkcolor=blue,urlcolor=blue}
\usepackage{color}
\usepackage{dcolumn}
\usepackage{lipsum}

\begin{document}

\title{Neutron occupancies and single-particle energies across the stable tin isotopes}

\author{S.~V.~Szwec} 
\altaffiliation[Present address: ]{University of Newcastle, University Drive, Callaghan 2308, NSW, Australia}
\affiliation{Department of Physics and Astronomy, University of Manchester, Manchester M13 9PL, United Kingdom}
\author{D.~K.~Sharp}
\email{david.sharp@manchester.ac.uk}
\affiliation{Department of Physics and Astronomy, University of Manchester, Manchester M13 9PL, United Kingdom}
\author{B.~P.~Kay}
\email{kay@anl.gov}
\affiliation{Physics Division, Argonne National Laboratory, Lemont, Illinois 60439, USA}
\author{S.~J.~Freeman}
\affiliation{Department of Physics and Astronomy, University of Manchester, Manchester M13 9PL, United Kingdom}
\author{J.~P.~Schiffer}
\affiliation{Physics Division, Argonne National Laboratory, Lemont, Illinois 60439, USA}
\author{P.~Adsley}
\altaffiliation[Present address: ]{University of the Witwatersrand, South Africa and iThemba LABS, South Africa}
\affiliation{Universit\'{e} Paris-Saclay, CNRS/IN2P3, IJCLab, 91405 Orsay, France}
\author{C.~Binnersley}
\affiliation{Department of Physics and Astronomy, University of Manchester, Manchester M13 9PL, United Kingdom}
\author{N.~de~S\'er\'eville}
\affiliation{Universit\'{e} Paris-Saclay, CNRS/IN2P3, IJCLab, 91405 Orsay, France}
\author{T.~Faestermann}
\affiliation{Physik Department E12, Technische Universit\"at M\"unchen, D-85748 Garching, Germany}
\affiliation{Maier-Leibnitz Laboratorium der M\"unchner Universit\"aten (MLL), D-85748 Garching, Germany}
\author{R.~F.~Garcia~Ruiz}
\altaffiliation[Present address: ]{Massachusetts Institute of Technology, Cambridge, Massachusetts 02139, USA}
\affiliation{Department of Physics and Astronomy, University of Manchester, Manchester M13 9PL, United Kingdom}
\author{F.~Hammache}
\affiliation{Universit\'{e} Paris-Saclay, CNRS/IN2P3, IJCLab, 91405 Orsay, France}
\author{R.~Hertenberger}
\affiliation{Maier-Leibnitz Laboratorium der M\"unchner Universit\"aten (MLL), D-85748 Garching, Germany}
\affiliation{Fakult\"at f\"ur Physik, Ludwig-Maximillians Universit\"at M\"unchen, D-85748 Garching, Germany}
\author{A.~Meyer}
\affiliation{Universit\'{e} Paris-Saclay, CNRS/IN2P3, IJCLab, 91405 Orsay, France}
\author{I.~Stefan}
\affiliation{Universit\'{e} Paris-Saclay, CNRS/IN2P3, IJCLab, 91405 Orsay, France}
\author{A.~Vernon}
\altaffiliation[Present address: ]{Massachusetts Institute of Technology, Cambridge, Massachusetts 02139, USA}
\affiliation{Department of Physics and Astronomy, University of Manchester, Manchester M13 9PL, United Kingdom}
\author{S.~Wilkins}
\altaffiliation[Present address: ]{Massachusetts Institute of Technology, Cambridge, Massachusetts 02139, USA}
\affiliation{Department of Physics and Astronomy, University of Manchester, Manchester M13 9PL, United Kingdom}
\author{H.-F.~Wirth}
\affiliation{Maier-Leibnitz Laboratorium der M\"unchner Universit\"aten (MLL), D-85748 Garching, Germany}
\affiliation{Fakult\"at f\"ur Physik, Ludwig-Maximillians Universit\"at M\"unchen, D-85748 Garching, Germany}


\begin{abstract}

The occupancies and vacancies of the valence neutron orbitals across the stable tin isotopic chain from $112\leq A\leq 124$ have been determined. These were inferred from the cross sections of neutron-adding and -removing reactions. In each case, the reactions were chosen to have good angular-momentum matching for transfer to the low- and high-$\ell$ orbitals present in this valence space. These new data are compared to older systematic studies. The effective single-neutron energies are determined by combining information from energy centroids determined from the adding and removing reactions. Two of the five orbitals are nearly degenerate, below $N=64$, and approximately two MeV more bound than the other three, which are also degenerate. 

\end{abstract}
\maketitle

\section{\label{intro}Introduction}

Knowledge of nuclear-structure properties of closed-shell nuclei and how they evolve with ($N-Z$) is essential to our understanding of the atomic nucleus~\cite{Sorlin08}. The exploration of how neutrons fill the $0g_{7/2}$, $1d_{5/2}$, $2s_{1/2}$, $1d_{3/2}$, and $0h_{11/2}$ orbitals along the tin isotopic chain from $112\leq A\leq124$, the longest chain of closed-shell, stable, even-$A$ isotopes, is the objective of this investigation. Such a systematic study was first attempted in the pioneering work of Cohen and Price~\cite{Cohen61}, who used the neutron-adding ($d$,$p$) and -removing ($d$,$t$) reactions on the stable, even-$A$ tin isotopes to determine the location of single-particle energies and the degree to which the orbitals were occupied.

Further systematic studies followed as reaction theory and experimental techniques developed, most notably those of Refs.~\cite{Schneid67,Cavanagh70,Fleming82}. There have been a number of studies on subsets of the stable tin isotopes using a wide variety of neutron-adding and -removing reactions over a broad range of beams energies, for example, Refs.~\cite{Nealy64,Cohen68,Yagi68,Bingham70,Mayer71,Bingham73,Bechara75,Borello-Lewin75,Vigdor75,Berrier-Ronsin77,Gerlic80,Blankert81,Dickey82,Gales82,Azaiez85,Massolo91,VandeWiele94} among others. Nucleon occupancies derived from the systematic studies are discrepant, often at the level of 20-40\%, and thus estimates of single-particle energies are uncertain by many 100s of keV. The goal of this work is to better constrain these properties.

\begin{figure}
\centering
\includegraphics[scale=0.8]{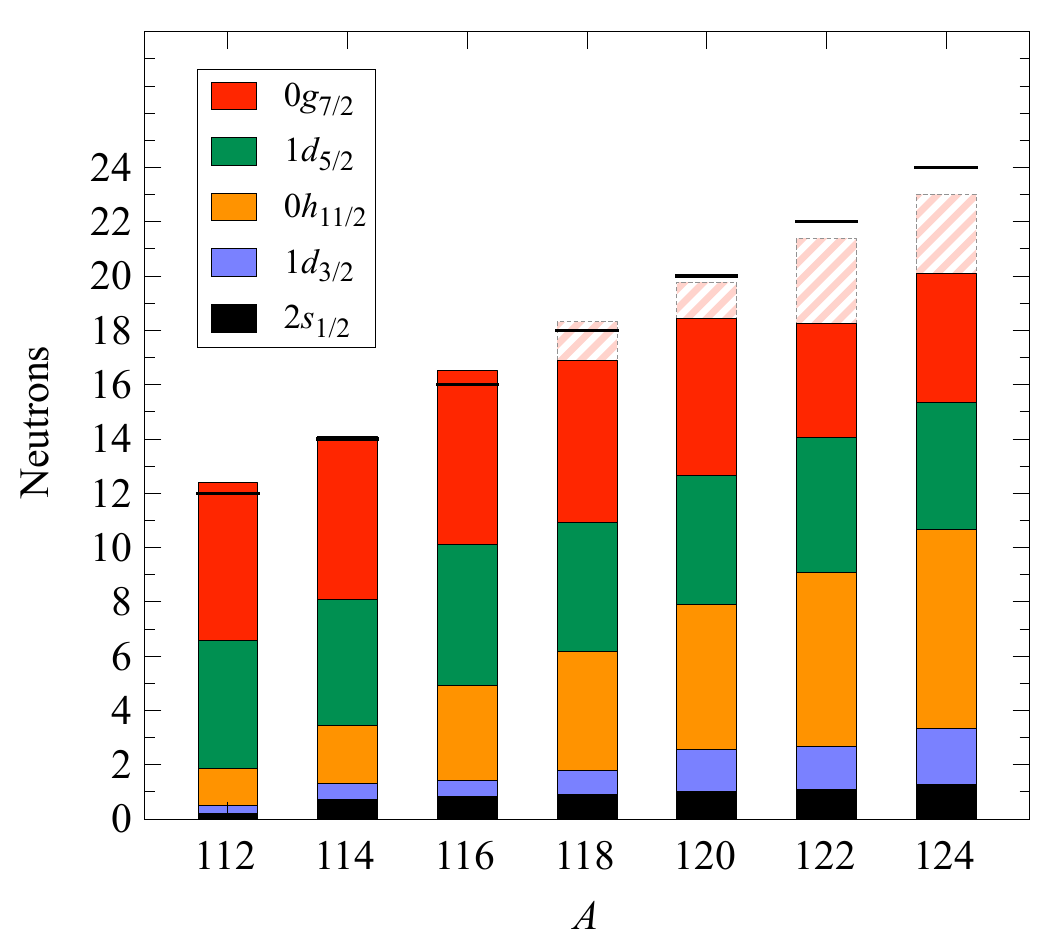}
\caption{\label{fig1} The filling of neutrons in the $0g_{7/2}$, $1d_{5/2}$, $2s_{1/2}$, $1d_{3/2}$, and $0h_{11/2}$ orbitals across the stable, even-$A$ tin isotopes as derived from this work. The horizontal bars are the nominal number of neutrons above $N=50$ for each isotope. The missing $0g_{7/2}$ strength (hatched) and the uncertainties are discussed in the text. }
\end{figure}

Figure~\ref{fig1} shows the filling of the neutron orbitals derived from the present study. A combination of different reactions were used to probe the single-neutron adding and removing strength. The result is a consistent description of filling of these orbitals to a level of uncertainty of a few tenths of a nucleon.

\begin{figure*}
\centering
\includegraphics[scale=0.8]{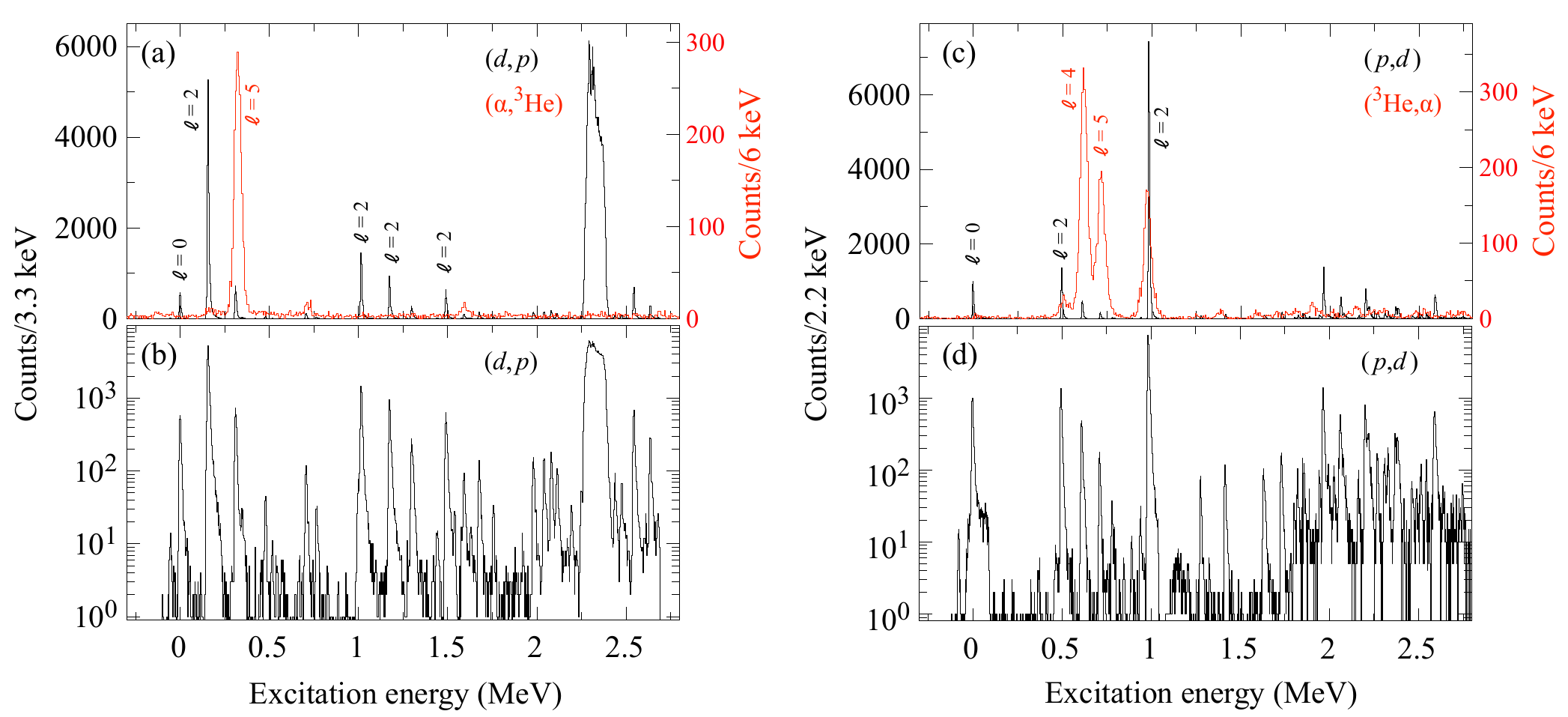}
\caption{\label{fig2} The $^{116}$Sn($d$,$p$)$^{117}$Sn and $^{116}$Sn($\alpha$,$^3$He)$^{117}$Sn reactions at 15~MeV ($\theta_{\rm lab}=18^{\circ}$) and 41~MeV ($\theta_{\rm lab}=10.9^{\circ}$), respectively, are shown in (a). Selected states are labeled by the transferred angular momentum $\ell$ to highlight the different matching conditions. Panel (b) is the same ($d$,$p$)-reaction data but with a logarithmic $y$-axis to emphasize the details of the spectrum. Similarly, the  $^{116}$Sn($p$,$d$)$^{115}$Sn and $^{116}$Sn($^3$He,$\alpha$)$^{115}$Sn  reactions at 21~MeV ($\theta_{\rm lab}=18^{\circ}$) and 36~MeV ($\theta_{\rm lab}=5.9^{\circ}$), respectively, in (c) and (d).  The broad peak around 2.3~MeV in Panels (a) and (b) is from reactions on the carbon target backing.}
\end{figure*}

This study follows recent work~\cite{Schiffer12} on the stable, \mbox{even-$A$} nickel isotopes which shows that the occupancies and how they change can be determined quantitatively, with well defined uncertainties, by paying particular attention to the choice of reactions, the experimental approach, and consistent analyses. Key to those studies were choosing reactions with good kinematic-matching conditions for transfer to orbits of different angular momenta, which is of particular importance in regions where both high- and low-$j$ orbitals are present, such as in the tin region.

To better determine the neutron occupancies, and thus single-neutron energies across the tin isotopes, new measurements of the ($p$,$d$), ($d$,$p$), ($^3$He,$\alpha$), ($\alpha$,$^3$He) reactions have been carried out. The magnitude of the $Q$ value for the ($p$,$d$) and ($d$,$p$) reactions are modestly low, around 3-8~MeV. The change in momentum between the incoming and outgoing ions for these reactions is well matched for transfer to final states reached by low-$\ell$, such as $1d_{5/2}$, and $2s_{1/2}$, $1d_{3/2}$, to within a unit or two of $\hbar$. In contrast, the magnitude of the $Q$ values for the ($^3$He,$\alpha$) and ($\alpha$,$^3$He) reactions are large, about 10-14~MeV, and well matched for $\ell=4$ and 5 transfer to the $0g_{7/2}$ and $0h_{11/2}$ orbitals. The striking impact of these matching conditions are shown in Fig.~\ref{fig2} for reactions on $^{116}$Sn, as studied in the current work. When carried out at energies a few MeV above the Coulomb barrier for the incoming and outgoing ions, the reaction cross sections are large, forward peaked, and can be reliably analyzed using the distorted-wave Born approximation (DWBA)~\cite{Schiffer12,Austern70}. 

\section{\label{exp}Experimental Methods}

\begin{table}
\caption{\label{tab1} Summary of the reactions and energies used on the even tin isotopes with $112\leq A\leq124$.}
\newcommand\T{\rule{0pt}{3ex}}
\newcommand \B{\rule[-1.8ex]{0pt}{0pt}}
\begin{ruledtabular}
\begin{tabular}{cccc}
Reactions & Lab & $E_{\rm beam}$~(MeV) & $\theta_{\rm lab}(^\circ)$  \\
\hline
\T($d$,$p$) & MLL & 15	& 6, 18, 30, 40 \\
($\alpha$,$^3$He) & IPN & 41 & 10.9 \\
($p$,$d$)& MLL & 21 & 6, 18, 30, 40 \\
($^3$He,$\alpha$) & IPN & 36 & 5.9\footnote{For $^{120,122}$Sn, additional angles of $\theta_{\rm lab}=10.9^{\circ}$, 15.9$^{\circ}$, and 20.9$^{\circ}$ were measured.}\\
 \end{tabular}
 \end{ruledtabular}
 \end{table}

The measurements presented in this paper were made at two different tandem accelerator facilities. The ($p$,$d$) and ($d$,$p$) reactions were measured at the Maier Leibnitz Laboratorium (MLL), taking advantage of the outstanding $Q$-value resolution of the Munich Q3D spectrometer. The ($^3$He,$\alpha$) and ($\alpha$,$^3$He) reactions were carried out at the Tandem-Alto facility at the Laboratoire de Physique des 2 Infinis Ir\`{e}ne Joliot-Curie (IJClab), where the tandem is capable of higher terminal voltages; necessary because of the large negative $Q$ values of the neutron-adding reaction. Table~\ref{tab1} is a summary of the reactions, targets, and angles, studied at each facility. Isotopically enriched targets of $^{112}$Sn (98.93\%), $^{114}$Sn (71.10\%), $^{116}$Sn (97.80\%),  $^{118}$Sn (98.60\%), $^{120}$Sn (99.70\%), $^{122}$Sn (96.00\%) and $^{124}$Sn (97.40\%) with a nominal thickness of $\sim$100$~\mu$g/cm$^2$ were used. The targets were evaporated onto a carbon backing of $\sim$20~$\mu$g/cm$^2$.

The instrumentation and methodology used for the MLL measurement was the same as that described in Ref.~\cite{Freeman17}. Similarly, Ref.~\cite{Szwec16} gives an overview of the approach taken at IJClab.

At MLL, the beams used were deuterons at 15~MeV and protons at 21~MeV at currents of 500-1000~nA. The outgoing ions from the reactions were momentum analyzed using the Q3D magnetic spectrograph~\cite{Scheerer76}. The entrance aperture of the spectrograph was fixed at values of 14.03~msr (nominal full aperture) or 7.25~msr (nominal half aperture) during the experiment. In order to extract absolute cross sections, measurements of the product of the target thickness and spectrometer entrance aperture for each target were made using elastic scattering of 10-MeV deuterons at $\theta_{\rm lab}=20^\circ$, integrated over the finite extent of the aperture. Under these conditions the measured cross section is estimated to be within 3\% of Rutherford scattering from optical-model calculations. The beam currents used for these calibrations were $\sim$1~nA, necessitating a different scale on the current integrator compared to the main reaction studies. The different integrator scales used were calibrated using a constant current source. 

\begin{figure}
\centering
\includegraphics[scale=0.75]{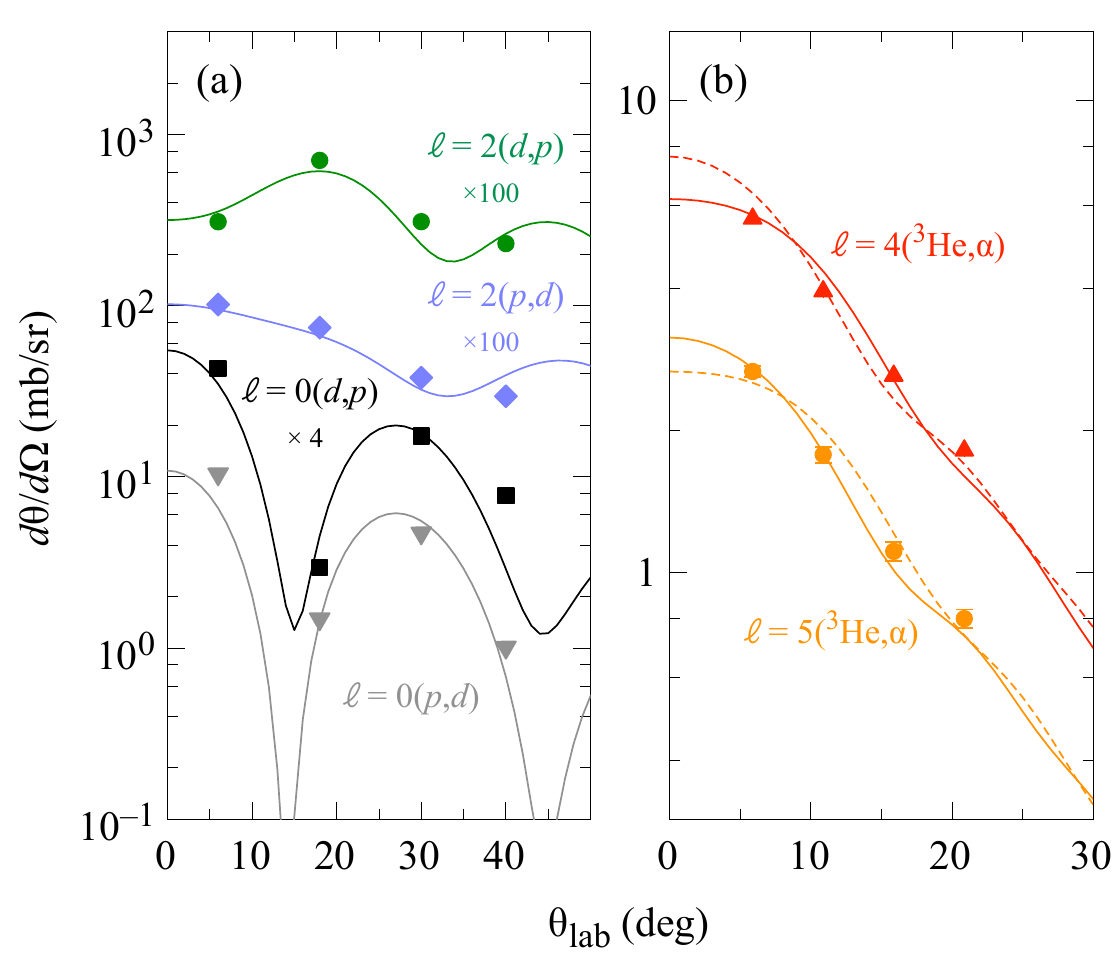}
\caption{\label{fig3} Example angular distributions for (a) $\ell=0$ and 2 transfer via the ($d$,$p$) and ($p$,$d$) reactions on $^{112}$Sn and $^{116}$Sn and (b) $\ell=4$ and 5 transfer via the ($^3$He,$\alpha$) reaction on $^{122}$Sn.}
\end{figure}

Cross sections were determined at the several angles selected to be at the first maxima of the angular distribution of the differential cross sections for the relevant $\ell$ transfers. First maxima were estimated by DWBA calculations to be at $\theta_{\rm lab}=6$, 18, 30 and 40$^{\circ}$ for $\ell=0$, 2, 4 and 5, respectively, in both the ($d$,$p$) and ($p$,$d$) reactions. Data taken at these angles also map out the angular distributions in a manner sufficient to determine the $\ell$ transferred in the reaction. While the majority of observed states have been identified in previous measurements, the angular distributions allow both confirmation of previous angular-momentum assignments and new $\ell$ assignments to be made. Reproduction of the measured angular distributions serves to validate choice of the parameterizations used in the DWBA, which is discussed in detail later. Figure~\ref{fig3} shows examples of angular distributions for each reaction, with a more complete record provided in the Supplemental Material~\cite{supmat}.

States in the residual nuclei were measured up to an excitation of $\sim$4~MeV in the ($d$,$p$) reaction and $\sim$3~MeV in the ($p$,$d$) reaction in most cases. For the $^{112,114}$Sn targets, the excitation-energy range for ($p$,$d$) reactions was limited due to the presence of elastically-scattered protons on the focal plane. Four magnetic-field settings were required in order to cover these ranges in excitation energy, with overlaps of several hundred keV. The states populated were calibrated in terms of excitation of the residual nucleus using states of known energy~\cite{nudat}. Typical spectra are shown in Fig.~\ref{fig2}; all spectra are available in the Supplemental Material~\cite{supmat}. The resolution obtained was $\sim$10~keV full-width half-maximum, and excitation energies were determined to better than 1~keV.

Reactions on oxygen and carbon, present in the targets, resulted in ions from reactions on these contaminants falling on the focal plane in the region of interest. These contaminants are easily identifiable due to their larger kinematic shift compared to the isotopes of interest, resulting in significant shifts in magnetic rigidity with angle and broader line shapes, as seen in the spectra of Fig.~\ref{fig2}(a-b). States corresponding to reactions on isotopic contaminants were also present and were identified by their rigidity. This was most apparent for reactions on $^{114}$Sn, which has a significantly lower isotopic purity than the other isotopes.

The ($\alpha$,$^3$He) and ($^3$He,$\alpha$) reactions were measured at IJClab. Beams of 36-MeV $^3$He and 41-MeV $\alpha$ particles at currents of 50-100~nA were used to bombard tin targets made from the same enriched material as used in the Munich experiments. Outgoing reaction products were momentum analyzed in an Enge split-pole spectrometer~\cite{Spencer67,Markham75}. The entrance aperture to the spectrometer was fixed for the duration of the experiment at a nominal value of 1.63~msr. The product of aperture and target thickness was measured using elastic scattering of $\alpha$ particles at an energy of 15~MeV and a laboratory angle of 20.9$^{\circ}$, which is in the Rutherford scattering regime. The full scale on the beam-current integrator was kept at a value of 10~nA for all measurements. The offset of the current source was calibrated using a known constant current source and found to be $<$1\%.

Angular distributions of the differential cross sections for both the ($\alpha$,$^3$He) and ($^3$He,$\alpha$) in this energy regime are forward peaked for the high-$\ell$ states of interest. Therefore, cross sections were measured at a nominal forward angles of 5.9$^{\circ}$---as far forward as is practical. For the ($\alpha$,$^3$He) reaction, this was limited to 10.9$^{\circ}$ due to other considerations. In order to assess the suitability of the choice of input into the DWBA reaction modeling, four-point angular distributions were measured for the ($^3$He, $\alpha$) reactions on targets of $^{120}$Sn and $^{122}$Sn, with 10.9$^{\circ}$, 15.9$^{\circ}$ and 20.9$^{\circ}$ being the additional angles. These are shown in Fig.~\ref{fig3}. The $\ell=4$ and 5 shapes are similar and discrimination between the two is not definitive. However, the spin-parity assignments for the vast majority of states populated are already known~\cite{nudat}.

The dispersion of the Enge split-pole spectrometer was such that the entire excitation-energy range of interest could be observed using a single magnetic-field setting. The observed resolution was 70~keV for $^3$He ions from the ($\alpha$,$^3$He) reactions and 90~keV for $^4$He ions from the ($^3$He,$\alpha$) reactions. Spectra are shown in Fig.~\ref{fig2} for these reactions on $^{116}$Sn. The energies of states were determined to better than 5-10~keV. All the states identified in these reactions were also observed in the low Q-value reactions.

Figures showing the spectra and tables of the measured cross sections on a state-by-state basis are given in the Supplemental Material~\cite{supmat}. The Supplemental Material also includes information on states populated in these reactions that correspond to orbitals outside of $50\leq N\leq82$, which are not the focus of the present study.

\section{\label{occ}Occupancies and vacancies}

Spectroscopic factors, $S_j$, are extracted through a comparison of the measured cross sections with those calculated using the DWBA approach. The exact finite-range DWBA code \textsc{ptolemy}~\cite{ptolemy} was used to calculate cross sections for the population of states in the residual nuclei. The spectroscopic factors were determined from the peak cross sections in the case of the ($p$,$d$) and ($d$,$p$) reactions for $\ell=2$ transfer, and for $\ell=0$ a fit to the secondary maxima as the forward angle was missing for some targets. Spectroscopic factors for the high angular-momentum states with $\ell=4$ and 5 were determined from ($\alpha$,$^3$He) and ($^3$He,$\alpha$) cross sections at $\theta_{\rm lab}=5.9^{\circ}$. For each type of reaction, for each state, a consistent approach to the reaction modeling is taken with the same sets of input parameterizations for the DWBA calculations for all the seven tin targets. The projectile bound-state parameters for the proton and deuteron are deduced using the Argonne $v_{18}$ potential~\cite{Wiringa95}. The $A=3$ and $A=4$ bound states are deduced via a Green's function Monte-Carlo method~\cite{Brida11}.

The bound-state potential describing the single-particle wave function of the odd neutron in the target nucleus used a fixed geometry of Woods-Saxon form with $r_0=1.28$~fm and $a_0=0.65$~fm. The depth of this potential is adjusted to reproduce the binding energies of the target. A spin-orbit component with depth $V_{\rm so}=6$~MeV, $r_{\rm so}=$1.10~fm and $a_{\rm so}=0.65$~fm is also included.

The optical potentials used to describe the incoming and outgoing distorted waves are, for the most part, taken from global parameterizations of elastic-scattering data. The deuteron optical-model-potential parameters are taken from the work of An and Cai~\cite{An06}, the protons from the work of Koning and Delaroche~\cite{Koning03} and the $^3$He parameters from Pang {\it et al.}~\cite{Pang09}. An optical-model parameter set fitted to data in the $A=90$ region~\cite{Bassani69} was used to describe the $\alpha$ potentials, which has been used for studies in the region before, for example on Sn~\cite{Schiffer04}.

The summed strength from both the adding ($G^+$) and removing ($G^-$) reactions to final states $i$ of a given single-particle orbital, $j$, should be equal to the degeneracy of that orbital such that
\begin{equation}\label{eqn1}
(2j+1)N_j=G^++G^-,~{\rm where}
\end{equation}
\begin{equation}\label{eqn2}
G^+=\sum_{i}(2j+1)C^2S^+_i~{\rm and}~G^-=\sum_{i}C^2S^-_i.
\end{equation}
Here, $S^+$ and $S^-$ are the spectroscopic factors for the removing and adding reactions. $C^2$ is the isospin Clebsch-Gordan coefficients~\cite{Schiffer69}, where $C^2=1$ for neutron adding, and with no proton occupancy of the relevant orbitals because of the closed proton shell, it does not enter into the removal reaction. $N_j$ is a normalization factor. This equation is an expression of the Macfarlane and French sum rules~\cite{Macfarlane60}. The normalization has been discussed extensively, but in most detail in Refs.~\cite{Schiffer12,Kay13,Aumann21}. 

The strength observed in single-nucleon transfer reactions at low momentum transfer has been shown to account for about ($N_j=$) 0.6(2) of the total single-particle strength, and appears to be independent of target mass, $\ell$ value, and reaction type~\cite{Kay13}, at least for nuclei near $\beta$ stability. To make an absolute determination of $N_j$ requires absolute cross sections. While absolute spectroscopic factors can vary by 20-30\% depending on choices in the experimental approach and reaction modeling, the variation in relative numbers obtained from this normalization procedure is less than 5\%.

\begin{figure}
\centering
\includegraphics[scale=0.78]{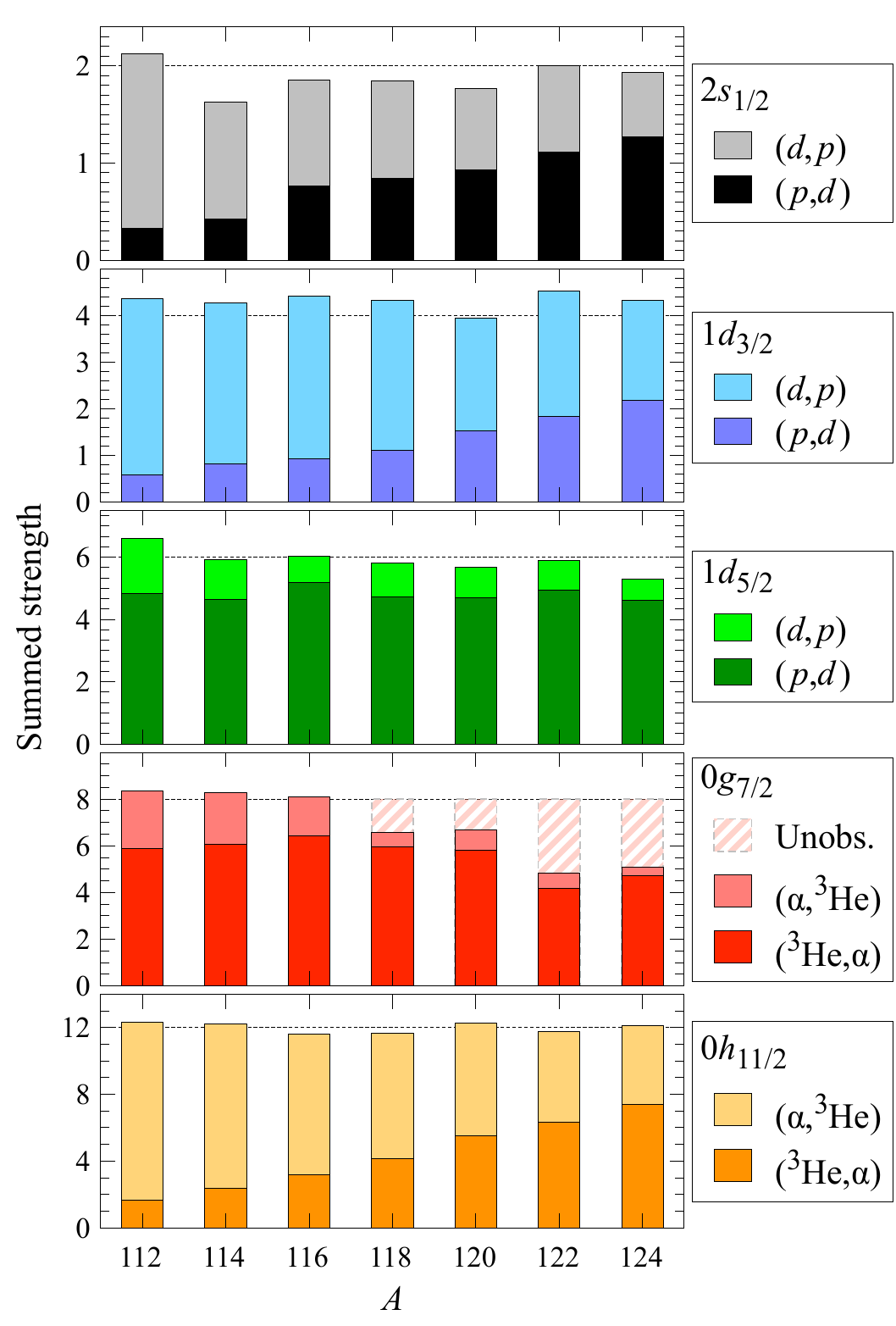}
\caption{\label{fig4} The summed strength for each orbital, highlighting the consistency. For the $0g_{7/2}$ orbital for $A>116$, there is reason to believe that strength at high excitation energies was outside the range studied here (see text for details), and this is emphasized by showing the missing portion as hatching.}
\end{figure}

Equation~\ref{eqn1} was used for each reaction populating each orbital across all seven targets. For $\ell=0$, 2 (the summed $1d_{5/2}$ and $1d_{3/2}$ was used in the normalization), 4, and 5, this produces seven values of $N_j$ (more strictly, $N_{\ell}$ in this case) each. To normalize the spectroscopic factors, a single average normalization $\overline{N_{02}}$ was used for the low-$Q$-value transfer, $\ell=0$, and 2. Over these fourteen values, $\overline{N_{02}}=1.07(9)$. The rms spread of around 10\% demonstrates remarkable consistency in the sum rules over the $2s_{1/2}$, $1d_{5/2}$, and $1d_{3/2}$ orbitals---this is highlighted in Fig.~\ref{fig4}. It suggests the essentially all of the low-lying $\ell=0$ and 2 strength was observed, consistently across the chain of isotopes, in these measurements. The value of the normalization appears higher than other similar studies. This could indicate some systematic shift in the absolute scale of the measured cross sections, but no other evidence for this was found. The consistency of the results across the Sn targets (discussed above) reveals no issue in the relative values of the ($p$,$d$) and ($d$,$p$) cross sections.

\begin{table}
\caption{\label{tab2} The summed neutron strength $G^{\pm}$ determined from the removing and adding reactions and the energy centroids, $E^{\pm}$, in keV. The uncertainties $G$ and $E$ are discussed in the text.}
\newcommand\T{\rule{0pt}{3ex}}
\newcommand \B{\rule[-1.8ex]{0pt}{0pt}}
\begin{ruledtabular}
\begin{tabular}{ccccc}
\B& $G^-_{2s_{1/2}}$ & $G^+_{2s_{1/2}}$ & $E^-_{2s_{1/2}}$ & $E^+_{2s_{1/2}}$  \\
\hline
\T$^{112}$Sn	&	0.33(3)	&	1.8(2)	&		507	&	441	\\
$^{114}$Sn	&	0.43(4)	&	1.2(1)	&		299	&	183	\\
$^{116}$Sn	&	0.76(7)	&	1.1(1)	&		222	&	130	\\
$^{118}$Sn	&	0.85(7)	&	1.0(1)	&		10	&	135	\\
$^{120}$Sn	&	0.93(8)	&	0.84(7)	&		161	&	81	\\
$^{122}$Sn	&	1.1(1)	&	0.89(8)	&		69	&	221	\\
$^{124}$Sn	&	1.3(1)	&	0.66(6)	&		237	&	215	\\
\hline
\T\B& $G^-_{1d_{3/2}}$ & $G^+_{1d_{3/2}}$ & $E^-_{1d_{3/2}}$ & $E^+_{1d_{3/2}}$ \\
\hline
\T$^{112}$Sn	&	0.59(1)	&	3.8(2)	&		654	&	510	\\
$^{114}$Sn	&	0.82(4)	&	3.4(2)	&		680	&	578	\\
$^{116}$Sn	&	0.93(4)	&	4.1(2)	&			673	&	159	\\
$^{118}$Sn	&	1.11(5)	&	3.2(2)	&		193	&	24	\\
$^{120}$Sn	&	1.5(1)	&	2.4(1)	&		24	&	121	\\
$^{122}$Sn	&	1.8(1)	&	2.7(1)	&		43	&	25	\\
$^{124}$Sn	&	2.2(1)	&	2.15(10)	&		25	&	28	\\
\hline
\T\B& $G^-_{1d_{5/2}}$ & $G^+_{1d_{5/2}}$ & $E^-_{1d_{5/2}}$ & $E^+_{1d_{5/2}}$  \\
\hline
\T$^{112}$Sn	&	4.8(3)	&	1.8(1)	&		226	&	630	\\
$^{114}$Sn	&	4.6(3)	&	1.3(1)	&		473	&	1271	\\
$^{116}$Sn	&	5.2(4)	&	0.86(6)	&		1041	&	1092	\\
$^{118}$Sn	&	4.7(3)	&	1.1(1)	&		1091	&	1108	\\
$^{120}$Sn	&	4.7(3)	&	0.98(7)	&		1143	&	1191	\\
$^{122}$Sn	&	4.9(4)	&	0.96(7)	&		1230	&	1290		\\
$^{124}$Sn	&	4.6(3)	&	0.69(5)	&		1330		&	1363	\\
\hline
\T\B& $G^-_{0g_{7/2}}$ & $G^+_{0g_{7/2}}$ & $E^-_{0g_{7/2}}$ & $E^+_{0g_{7/2}}$ \\
\hline
\T$^{112}$Sn	&	5.9(1)	&	2.46(4)	&		147	&	77	\\
$^{114}$Sn	&	6.1(1)	&	2.22(4)	&	 		151	&	944	\\
$^{116}$Sn	&	6.4(1)	&	1.65(3)	&		671	&	712	\\
$^{118}$Sn	&	7.1$^{+0.3}_{-0.1}$	&	0.6(1)	&		712	&	787	\\
$^{120}$Sn	&	6.9$^{+0.2}_{-0.1}$	&	0.9(2)	&		787	&	926	\\
$^{122}$Sn	&	6.7$^{+0.6}_{-0.1}$	&	0.7(1)	&		926	&	1044	\\
$^{124}$Sn	&	4.7$^{+3.0}_{-0.1}$ &	0.3(1)	&		1135	&	1363	\\
\hline
\T\B& $G^-_{0h_{11/2}}$ & $G^+_{0h_{11/2}}$ & $E^-_{0h_{11/2}}$ & $E^+_{0h_{11/2}}$ \\
\hline
\T$^{112}$Sn	&	1.66(3)	&	10.7(2)	&		979	&	767	\\
$^{114}$Sn	&	2.36(4)	&	9.8(2)	&		738	&	815	\\
$^{116}$Sn	&	3.19(6)	&	8.4(2)	&		858	&	315	\\
$^{118}$Sn	&	4.14(7)	&	7.5(1)	&		315	&	90	\\
$^{120}$Sn	&	5.5(1)	&	6.8(1)	&		90	&	6	\\
$^{122}$Sn	&	6.3(1)	&	5.5(1)	&		6	&	0	\\
$^{124}$Sn	&	7.4(1)	&	4.7(1)	&		0	&	0	\\
 \end{tabular}
 \end{ruledtabular}
 \end{table}

The summed $\ell=4$ strength determined from the ($\alpha$,$^3$He) and ($^3$He,$\alpha$) reaction is approximately constant for $^{112}$Sn, $^{114}$Sn, and $^{116}$Sn, but drops beyond that to about 60\% by $^{124}$Sn. This is indicative of missing strength, and most likely dominantly in the neutron-removal reactions. As neutron number increases across the tin isotopes the $g_{7/2}$ orbital becomes more deeply bound and the associated single-particle strength shifts to higher excitation energy in the residual nuclei and becomes more fragmented. Such behavior is to be expected~\cite{Kay13a,Szwec16} and can be seen in Fig.~\ref{fig4}, with the observed strength decreasing for $A>116$. In order to obtain a normalization for the high angular momentum states with $\ell=4$ and 5, the summed $0g_{7/2}$ strength for $^{112,114,116}$Sn and all targets for the $0h_{11/2}$ strength was used, yielding $\overline{N_{45}}=0.71(4)$. If just the $0h_{11/2}$ strength was used, $\overline{N_{5}}=0.69(3)$. The consistency of $0h_{11/2}$ strength, which is dominated by a single strong state for both the adding and removing reactions on each target, supports the conclusions that there is unobserved $0g_{7/2}$ strength in the removing reactions. Like the low-$\ell$ transfer, probing the $2s_{1/2}$ and $1d$ strength, the high-$\ell$ data yields remarkable consistency in the sum rules across the tin isotopes. It is noted that for the $0g_{7/2}$ and $0h_{11/2}$, $\overline{N}$ is consistent with similar data presented in Ref.~\cite{Kay13}.

The deduced adding and removing strength, along with the summed strength is given in Tab.~\ref{tab2}, along with the centroids of single-particle strength as determined from the adding and removing reactions, $E^{\pm}$. This is defined as the spectroscopic-factor weighted energy
\begin{equation}\label{eqn3}
 E_j^{\pm}=\sum_i E^{*\pm}_j(i)S_j(i)/\sum_i S_j(i),
\end{equation}
where $E^{*}$ is the energy of a given excited state $i$ in an orbital $j$. The centroids are used in the determination of effective single-particle energies in the following section. The uncertainties, $dE^{\pm}$, in the centroids are estimated to be $-30\lesssim dE\lesssim +150$~keV, except in cases where it the strength is dominated by the ground state or a single low low-lying state, then the lower limit has an uncertainty of typically a keV or so. The upper limit is a conservative estimate, taking into account possible unobserved strength. The uncertainties are larger for the $0g_{7/2}$ orbital for $^{118-124}$Sn, estimated to be around 300~keV.
 
\begin{figure}
\centering
\includegraphics[scale=0.75]{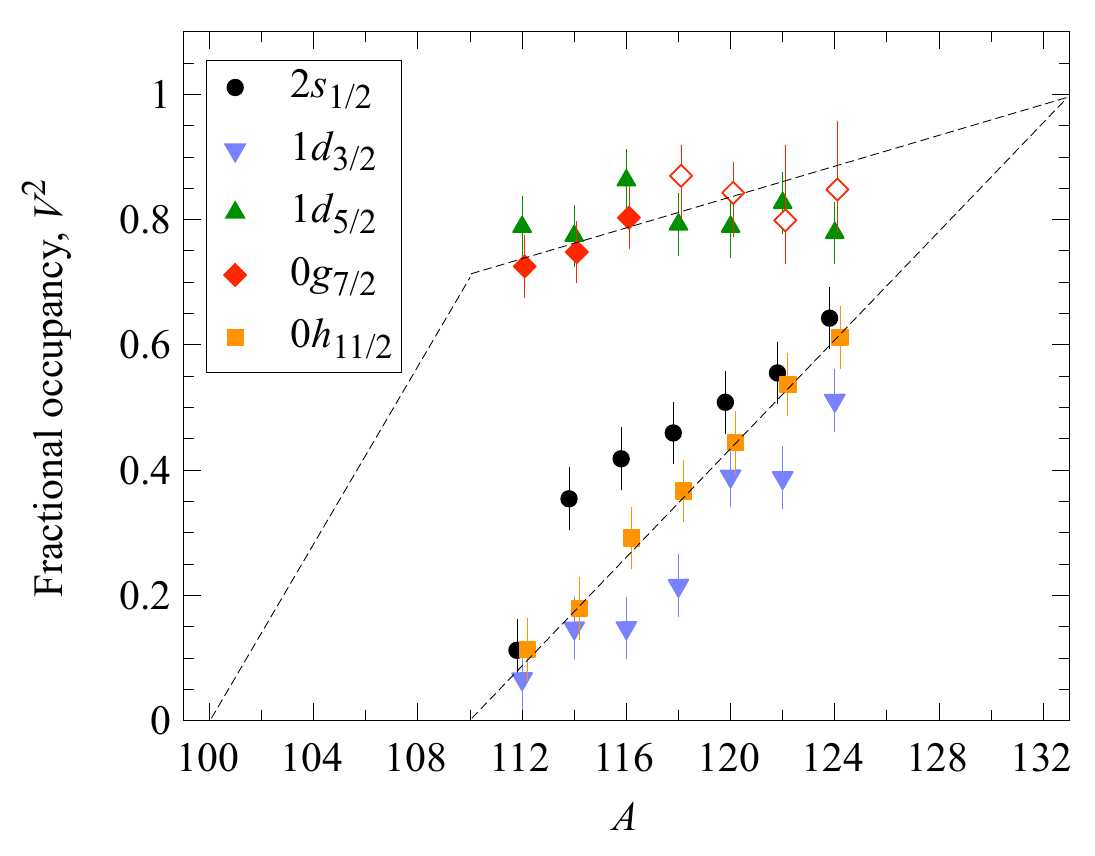}
\caption{\label{fig5} The fractional occupancy of the neutron $0g_{7/2}$, $1d_{5/2}$, $2s_{1/2}$, $1d_{3/2}$, and $0h_{11/2}$ orbitals as deduced in this work. The dashed lines are to guide the eye and the $0g_{7/2}$ data with open symbols have been adjusted using an approximation discussed in the text.}
\end{figure}

The rms spread can provide a measure of the uncertainties in the summed strengths, the consistency in the normalizations for different orbitals adding up to $2J+1$ for each isotope. The internal consistency of the data and the normalization procedure work to a level better than a few percent. The principal uncertainties in the sums are from possible missed strength, as evidenced by the $\ell=4$ missing strength already discussed. Another source of uncertainty are missing assignments of $j^{\pi}$, which are most apparent for $\ell=2$ transfer where in several cases no firm assignment of 3/2$^+$ and 5/2$^+$ can be made. The unassigned levels represent less than 10\% of the total across the targets studied here. The uncertainty in the counting statistics are generally small, being less than 1\% for weak fragments.  The overall uncertainties in the summed strength are estimated to be between 0.1-0.9 nucleons, guided largely by the rms spread in the normalization, as shown in Tab.~\ref{tab2}. This corresponds to an uncertainty in the fractional occupancy of around 0.05.

The fractional neutron occupancies, $V^2$, given in Tab.~\ref{tab3} and shown in Fig.~\ref{fig5} are derived from the weighted average of the adding and removing strength. The $0g_{7/2}$ fractional occupancies presented Tab.~\ref{tab3} and Fig.~\ref{fig5} have been adjusted $^{118\mbox{-}124}$Sn, using the somewhat arbitrary assumption that 70\% of the missing strength discussed above is in the removal reaction. The error bars account for this adjustment.

\begin{table}
\caption{\label{tab3} Fractional occupancy, $V^2$. Uncertainties are discussed in the text.}
\newcommand\T{\rule{0pt}{3ex}}
\newcommand \B{\rule[-1.8ex]{0pt}{0pt}}
\begin{ruledtabular}
\begin{tabular}{cccccc}
\B&$2s_{1/2}$ & $1d_{3/2}$ & $1d_{5/2}$ & $0g_{7/2}$\footnote{The magnitude of the fractional occupancies for $^{118\mbox{-}124}$Sn are discussed in the text.} & $0h_{11/2}$\\
\hline
\T$^{112}$Sn	&	0.11(5)	&	0.07(5)	&	0.79(5)	&	0.73(5)	&	0.11(5)	\\
\T$^{114}$Sn	&	0.35(5)	&	0.15(5)	&	0.77(5)	&	0.75(5)	&	0.18(5)	\\
\T$^{116}$Sn	&	0.42(5)	&	0.15(5)	&	0.86(5)	&	0.80(5)	&	0.29(5)	\\
\T$^{118}$Sn	&	0.46(5)	&	0.22(5)	&	0.79(5)	&	0.87$^{+0.05}_{-0.07}$	&	0.37(5)	\\
\T$^{120}$Sn	&	0.51(5)	&	0.39(5)	&	0.79(5)	&	0.84$^{+0.05}_{-0.07}$	&	0.44(5)	\\
\T$^{122}$Sn	&	0.56(5)	&	0.39(5)	&	0.83(5)	&	0.80$^{+0.12}_{-0.07}$	&	0.54(5)	\\
\T\B$^{124}$Sn	&	0.64(5)	&	0.51(5)	&	0.78(5)	&	0.85$^{+0.11}_{-0.07}$	&	0.61(5)	\\
 \end{tabular}
 \end{ruledtabular}
 \end{table}

\begin{figure}
\centering
\includegraphics[scale=0.75]{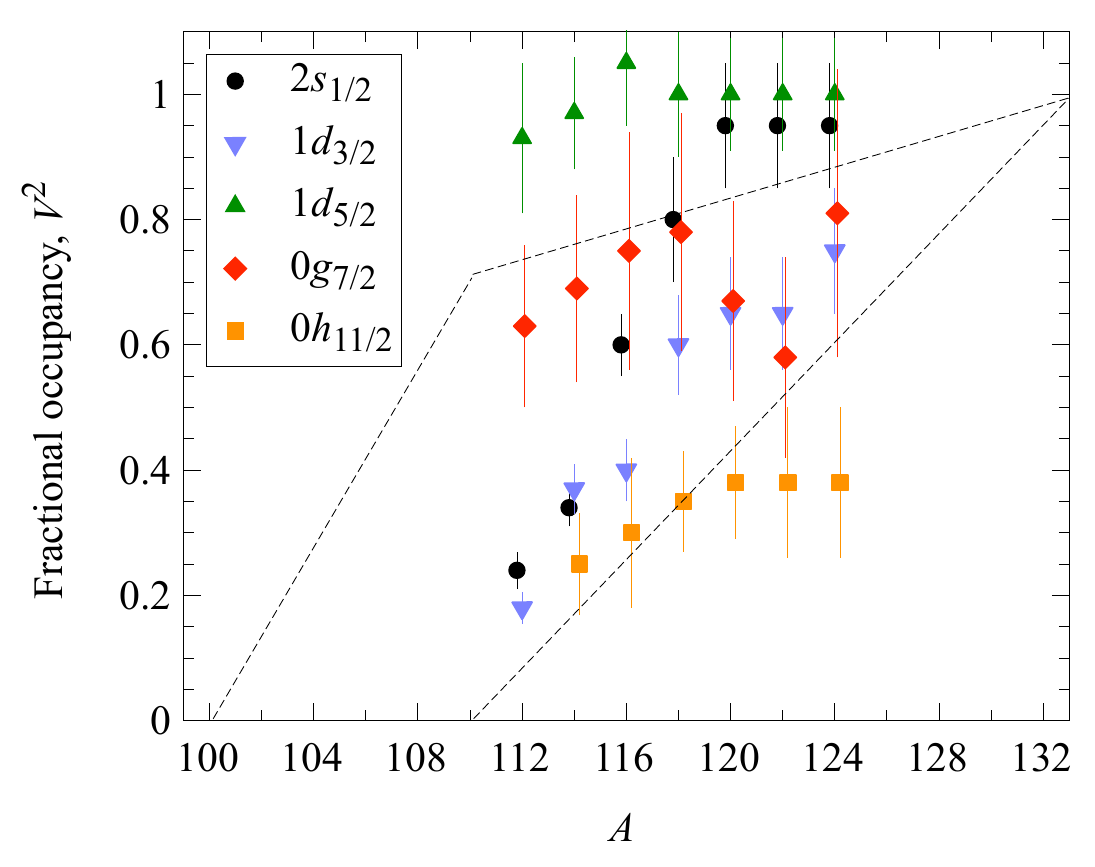}
\caption{\label{fig6} The fractional occupancy of the neutron $0g_{7/2}$, $1d_{5/2}$, $2s_{1/2}$, $1d_{3/2}$, and $0h_{11/2}$ orbitals as deduced in the study of Ref.~\cite{Fleming82}. The dashed lines are are the same as in Fig.~\ref{fig5}.}
\end{figure}

The data shown in Fig~\ref{fig6} are the fractional occupancies derived in the study of Fleming~\cite{Fleming82} (summarized in Table 5 of that reference), which was a systematic exploration of the neutron occupancy using the ($p$,$d$) reaction. Qualitatively, a similar pattern is seen, with the $1d_{5/2}$ and $0g_{7/2}$ orbitals filled at an approximately constant level across the isotopic chain, and the $2s_{1/2}$, $1d_{3/2}$, and $0h_{11/2}$ orbitals filling across the chain. The details are quite different though. The new data reveal that the $2s_{1/2}$, $1d_{3/2}$, and $0h_{11/2}$ fill at a similar manner, and reaching only about 60\% occupancy by $^{124}$Sn, in contrast to the previous study where they fill at markedly different rates, with the $2s_{1/2}$ orbital essentially full at $^{124}$Sn, the $1d_{3/2}$ around 70\% and the $0h_{11/2}$ only 30\% full. The lower $\ell=5$ strength determined from just the ($p$,$d$) reaction is possibly a consequence of the less optimal matching conditions. The missing $\ell=4$ strength for $A>116$ is also apparent in the ($p$,$d$) reaction data of Fleming.

\begin{figure}
\centering
\includegraphics[scale=0.75]{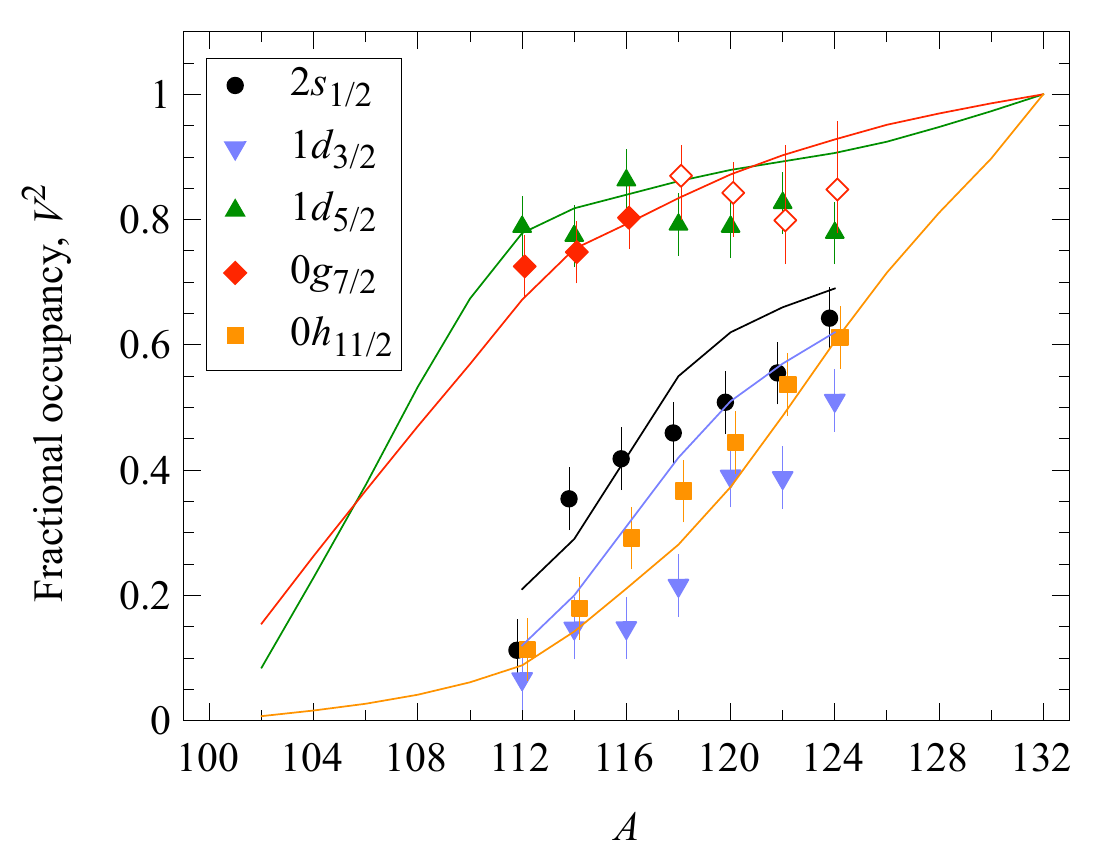}
\caption{\label{fig7} A comparison of the fractional occupancies determined in this work compared with shell-model calculations from Ref.~\cite{Qi12}, for which information is available for the $1d_{5/2}$, $0g_{7/2}$, and $0h_{11/2}$ orbitals for $102\leq A\leq132$, and for the $2s_{1/2}$ and $1d_{3/2}$, $112\leq A\leq124$ only.}
\end{figure}

Figure~\ref{fig7} compares the fractional occupancies derived from this study to those determined using shell-model calculations as presented in Ref.~\cite{Qi12}, with a new interaction that used the CD-Bonn nucleon-nucleon force as a starting point. Fractional occupancies were calculated for $1d_{5/2}$, $0g_{7/2}$, and $0h_{11/2}$ orbitals for $102\leq A\leq132$, and for the $2s_{1/2}$ and $1d_{3/2}$, $112\leq A\leq124$ only. The theoretical data compares favorably to the experimental data, with the $1d_{5/2}$, $0g_{7/2}$, and $0h_{11/2}$ orbitals being well described by theory. The largest discrepancy is in the description of the filling of the $1d_{3/2}$ orbital, which fills more slowly than the theory predicts, and conversely, the $2s_{1/2}$ orbital fills less slowly.

The trends shown in Fig.~\ref{fig7} raises the question of how the $1d_{5/2}$ and $0g_{7/2}$ orbitals fill between $100\lesssim A\lesssim 112$, where they both seem to be about 80\% from 112 to 124. The theoretical calculations suggest they fill in parallel at a similar rate, much like the $2s_{1/2}$, $1d_{3/2}$, and $0g_{7/2}$ orbitals do starting around $A\sim110$, but there is essentially no information on the fractional occupancies for these unstable nuclei. Targeted measurements would be of interest, and are soon to be possible at the next generation of radioactive ion beam facilities coming online.

\section{\label{spes}Single-particle energies}

The effective single-particle energies, $\epsilon_j$, can be determined from the centroids of single-particle strength for the adding and removing reactions, that are derived as described in Eq.~\ref{eqn3}. The effective single-particle energy was formally defined by Baranger~\cite{Baranger70}, and is equivalent in definition to the monopole formulation discussed in the recent review article of Otsuka {\it et al.}~\cite{Otsuka20}, which has been used extensively in describing the evolution of effective single-particle energies with neutron excess. 

The effective single-particle energies, $\epsilon_j$, are defined as the combination of the centroids of single-particle strength, $E^{\pm}$, (Eq.~\ref{eqn3}), and on an absolute scale with respect to zero binding, as:
\begin{equation}\label{eqn4}
\epsilon_j=\frac{E_j'^+G_j^++E_j'^-G_j^-}{G_j^++G_j^-}
\end{equation}
where $G^{\pm}$ are the normalized summed strengths of Tab.~\ref{tab2} and
\begin{equation}\label{eqn5}
E_j'^+=-B(A+1)+E_j^+,
\end{equation}
with $B$ being the binding energy of the target plus a neutron system in the adding reaction and $E^+$ the centroid of Eq.~\ref{eqn3}. Similarly,
\begin{equation}\label{eqn6}
E_j'^-=-B(A)-E_j^-.
\end{equation}

\begin{figure}
\centering
\includegraphics[scale=0.78]{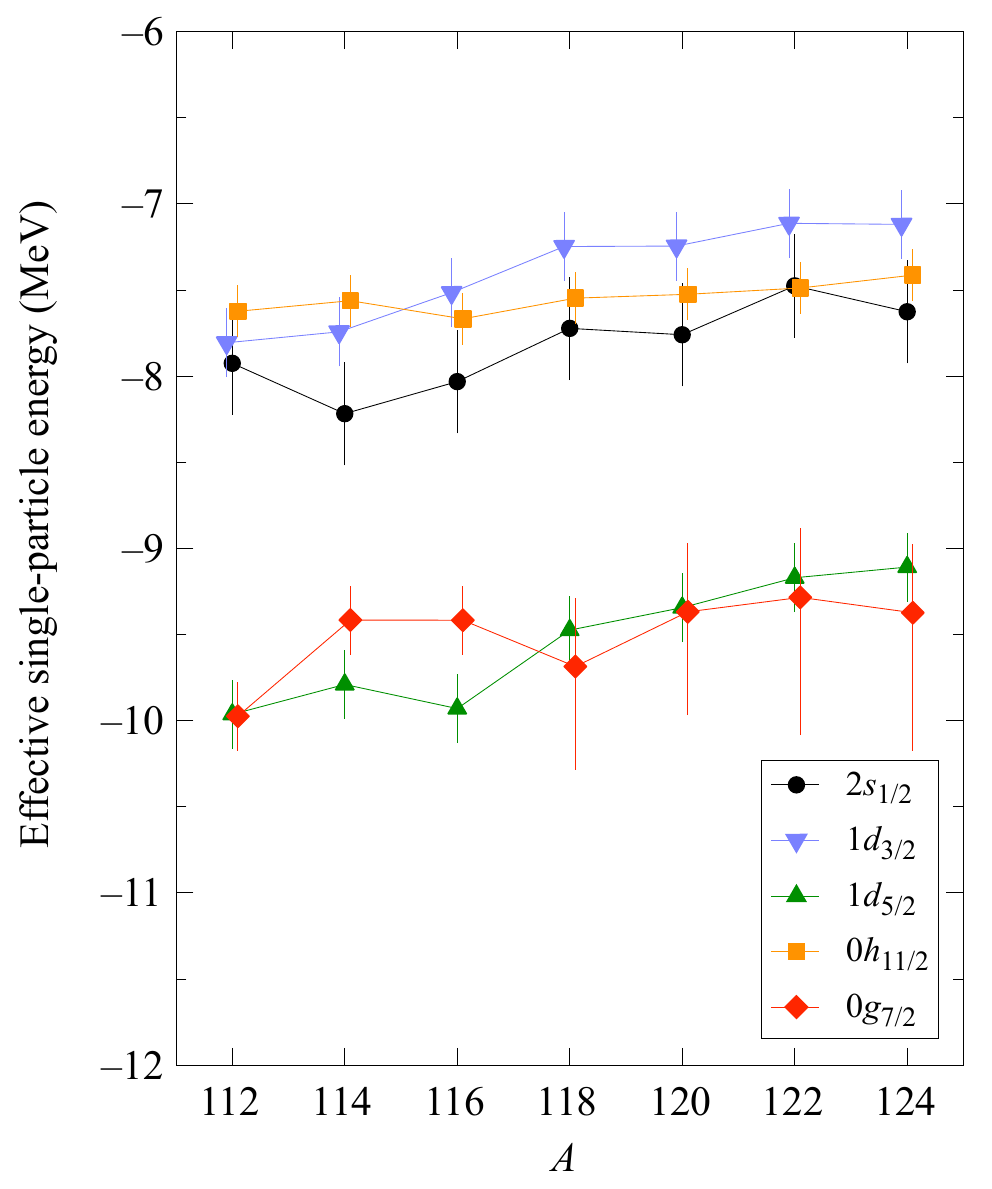}
\caption{\label{fig8} The effective single-particle energies for neutron orbitals across the stable, even-$A$ tin isotopes.}
\end{figure}

\begin{figure}
\centering
\includegraphics[scale=0.75]{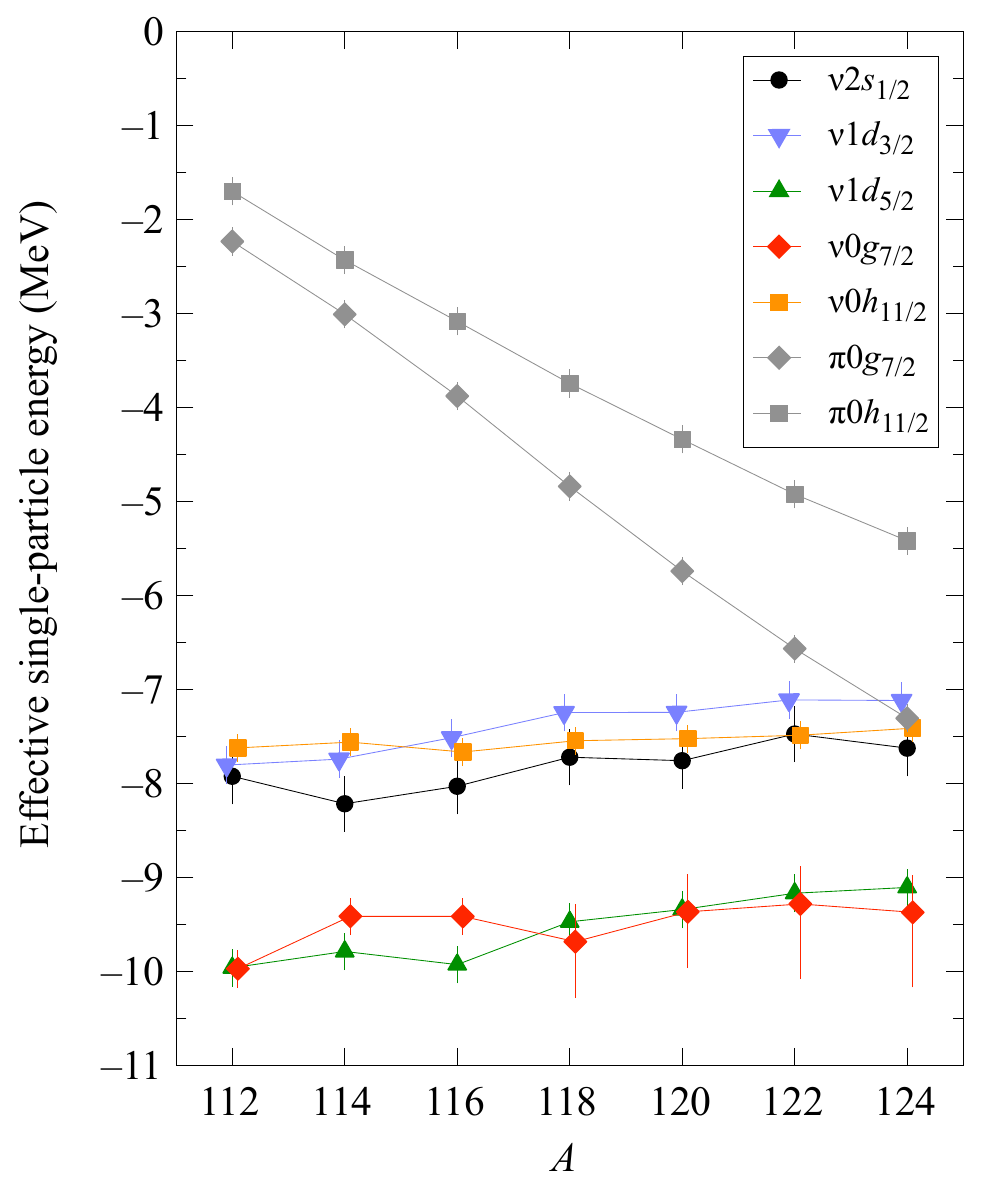}
\caption{\label{fig9} The effective single-particle energies for neutron orbitals across the stable, even-$A$ tin isotopes contrasted with those of protons for the $0g_{7/2}$ and $0h_{11/2}$ orbitals, which are well determined from the ($\alpha$,$t$) reaction~\cite{Schiffer04}. }
\end{figure}

Figure~\ref{fig8} shows the effective single-neutron energies as determined using Eq.~\ref{eqn4} for $0g_{7/2}$, $1d_{5/2}$, $2s_{1/2}$, $1d_{3/2}$, and $0h_{11/2}$ orbitals. Numerical values are given in Tab.~\ref{tab4}. The binding energy of all of these orbitals deceases slowly across the tin isotopes, each at essentially the same rate of around 50-100~keV per additional neutron. By contrast, the proton orbitals' binding energies change by $\sim$350-400~keV per additional neutron over the same range~\cite{Schiffer04} as shown in Fig.~\ref{fig9}. The contrast is striking---the neutron single-particle states become less bound slowly, while the proton single particle states become much more bound with increasing neutron number. This pattern is being studied more broadly~\cite{tobepublished}. 

\begin{table}
\caption{\label{tab4} Effective single-particle energies in MeV. Uncertainties are discussed in the text.}
\newcommand\T{\rule{0pt}{3ex}}
\newcommand \B{\rule[-1.8ex]{0pt}{0pt}}
\begin{ruledtabular}
\begin{tabular}{cccccc}
\B&$2s_{1/2}$ & $1d_{3/2}$ & $1d_{3/2}$ & $0g_{7/2}$ & $0h_{11/2}$ \\
\hline
\T$^{112}$Sn	&	$-$7.9(3)	&	$-$7.8(2)	&	$-$10.0(2)	&	$-$10.0(2)	&	$-$7.62(15)	\\
\T$^{114}$Sn	&	$-$8.2(3)	&	$-$7.7(2)  &	$-$9.8(2)	&	$-$9.4(2)	&	$-$7.56(15)	\\
\T$^{116}$Sn	&	$-$8.0(3)	&	$-$7.5(2)	&	$-$9.9(2)	&	$-$9.4(2)	&	$-$7.67(15)	\\
\T$^{118}$Sn	&	$-$7.7(3)	&	$-$7.2(2)	&	$-$9.5(2)	&	$-$9.7$^{+0.4}_{-0.6}$	&	$-$7.55(15)	\\
\T$^{120}$Sn	&	$-$7.8(3)	&	$-$7.2(2)	&	$-$9.3(2)	&	$-$9.4$^{+0.4}_{-0.6}$	&	$-$7.52(15)	\\
\T$^{122}$Sn	&	$-$7.5(3)	&	$-$7.1(2)	&	$-$9.2(2)	&	$-$9.2$^{+0.4}_{-0.8}$	&	$-$7.49(15)	\\
\T\B$^{124}$Sn	&	$-$7.6(3)	&	$-$7.1(2)	&	$-$9.1(2)	&	$-$9.4$^{+0.4}_{-0.8}$	&	$-$7.41(15)	\\
 \end{tabular}
 \end{ruledtabular}
 \end{table}

Another striking feature is the near degeneracy of the $2s_{1/2}$, $1d_{3/2}$, and $0h_{11/2}$ orbitals, which fill in parallel across the isotopic chain, and similarly that of the other two orbitals $1d_{5/2}$ and $0g_{7/2}$, that are already mostly filled across the same region. The two degenerate groupings are well separated by around 2~MeV, similar in magnitude to the pairing gap. This suggests strong correlations within each of the two sets of degenerate orbitals, and some, considerably weaker, correlation between the two.

The uncertainties on the effective single-particle energies reflect those in the summed strength from adding and removing reactions, and the same considerations with regards to reaction modeling, unassigned and mis-assigned strength. We note that the magnitude of the normalization factor, $N_j$, has no impact on the single-particle energy. Only changes in the centroids and thus the distribution of strength in adding and removing have an effect. For the $2s_{1/2}$ and $1d$ orbitals, variations in either the summed adding or removing strength by 10\%, or both, which is commensurate with the rms spread of the total strength, and possible unobserved strength, result in a estimated uncertainty in the effective single-particle energies by around $\pm$300~keV for the $2s_{1/2}$ orbital and $\pm$200~keV for the $1d$ orbitals. The rms spread in the summed $0h_{11/2}$ strengths is smaller, around 5\%, resulting in an estimated $\pm$150~keV variation in the single-particle energy. As discussed above, there is robust evidence of unobserved $0g_{7/2}$ strength, dominantly in the neutron-removal reaction, which results in asymmetric uncertainties in the single-particle energies, where the orbital is likely more bound due to this unobserved strength. For $^{118,120}$Sn, this results in effective single-particle energies of $+400<\Delta\epsilon_{0g_{7/2}}<-600$~keV and for $^{122,124}$Sn,  $+400<\Delta\epsilon_{0g_{7/2}}<-800$~keV. We note that the adjustment to the $0g_{7/2}$ fractional occupancy was not used in the determination of the effective single-particle energies, and this is reflected in the uncertainties.

\section{\label{BCS}Analysis of results in the BCS framework}

\begin{figure}
\centering
\includegraphics[scale=0.68]{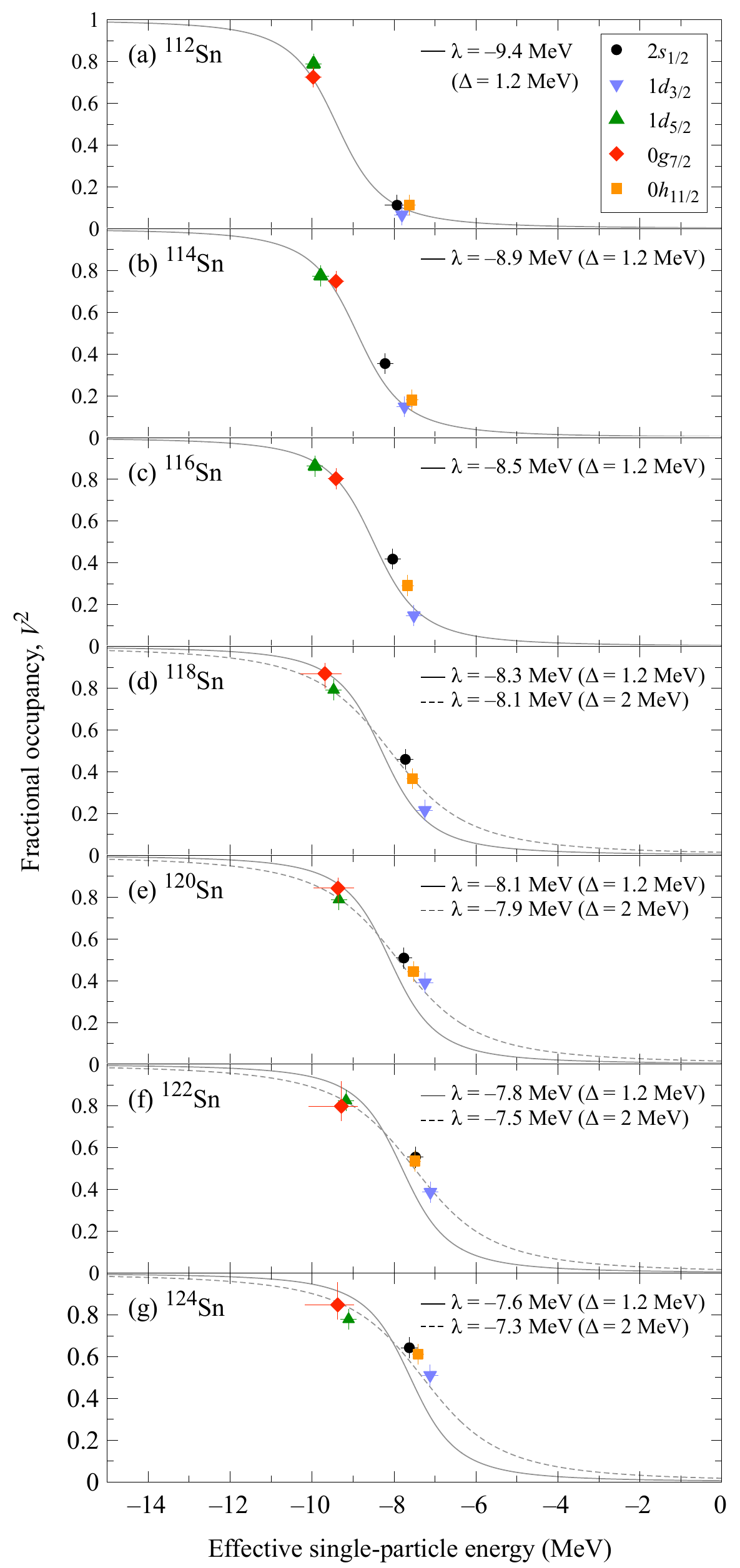}
\caption{\label{fig10} The distribution of the effective single-neutron energies about the Fermi surface for $^{112-124}$Sn (a-g) as a function of fractional occupancy, $V^2$. The solid curves are the BCS occupation probabilities calculated with parameters of $\lambda$ varying smoothly from $-$9.4~MeV to $-$7.3~MeV across the range and $\Delta=1.2$~MeV, as defined in the text, aside from the dashed lines (d-g) which are for $\Delta=2$~MeV.}
\end{figure}

The simple pairing approximation, based on the BCS concept~\cite{Bardeen57,Kisslinger63}, can relate the occupation numbers, $U^2$ and $V^2$ (emptiness and fullness, such that $U^2+V^2=1$) to the effective single-particle energies, the Fermi energy $\lambda$, and the pairing gap $\Delta$. In the early transfer reaction studies on the Sn isotopes by the Pittsburg group~\cite{Cohen61,Schneid67}, the occupation numbers extracted from those works were compared to pairing theory, though discrepancies (sometimes by as much as an MeV) were noted when comparing data to the centroids of single-particle strength (effective single-particle energies were not discussed). At least a part of these discrepancies may be attributed to missed states and spin-parity assignments that were not known at the time. 

In Fig.~\ref{fig10}, the occupation probability (or fractional occupancy, $V^2$) is plotted as a function of effective single-particle energy and compared against the BCS function, defined as
\begin{equation}\label{eqn}
{V^2}=\frac{1}{2}\left[1-\frac{(\epsilon_j-\lambda)}{[(\epsilon_j-\lambda)^2+\Delta^2]^{1/2}}\right].
\end{equation}
From the data, $\Delta$ was estimated to be around 1.2~MeV (similar to the nominal 1.3-1.4~MeV determined from the masses), but for the heavier tin isotopes, it appears a larger value of $\Delta$ is needed, suggesting a more diffuse Fermi surface. The value of $\lambda$ was inferred by fitting to the data. The Fermi surface gradually increases from around $-$9.4~MeV to $-$7.3~MeV across the chain of isotopes as the $2s_{1/2}$, $1d_{5/2}$, and $0h_{11/2}$ orbitals fill essentially in parallel. As noted in previous works there are discrepancies between the experimental data and BCS theory, but these are removed by a modest increase in the pairing gap. Possibly the BCS mixing of states is first between the $0g_{7/2}$ and $1d_{5/2}$ states, then a larger gap at $N=64$, and then the mixing between the $2s_{1/2}$, $1d_{3/2}$, and $0h_{11/2}$ orbitals that are almost degenerate.

The data suggest correlations between all five orbits, but with some apparent substructure between the two sets of degenerate orbitals as discussed above. Enhanced cross sections in two-neutron transfer ($t$,$p$) or ($p$,$t$) reactions are the hallmark of BCS correlations in nuclear ground states. In one of the studies of ($p$,$t$) reactions on the even Sn isotopes there seems to be little evidence for significant population of an excited 0$^+$ state, suggesting that the correlation characteristics of this picture are concentrated in the ground states (see for instance Ref.~\cite{Fleming70}, where the evidence is summarized in the discussion section of that paper). There is little evidence of a `pairing vibration,’ that would be present if the $1d_{5/2}$ and $0g_{7/2}$ orbits were mixing only with each other, and did not mix into the BCS condensate with the other three orbits in this region. The results on occupancies presented here, appear to be qualitatively consistent with the data from two-nucleon transfer that the BCS condensate for neutrons in the Sn isotopes involves all five orbits between $N=50$ and 82, but perhaps with some minor modifications.

\section{\label{summary}Summary}

Single-nucleon occupancies and effective single-particle energies for the valence neutron orbitals in the stable, even-$A$ Sn isotopes have been determined. The results show remarkable consistency between the adding and removing reactions. Of the five orbitals in this region, three are found to be closely correlated in the degree of filling and virtually degenerate in their effective single-particle energies. The other two are mostly filled in the stable tin region, but their energies are also almost degenerate.

\section{acknowledgments}

We would like to acknowledge the accelerator operating staff and target makers at the Maier-Leibnitz Laboratorium der M\"unchner Universit\"aten as well as the operating staff at the Laboratoire de Physique des 2 Infinis Ir\`{e}ne Joliot-Curie. This material is based upon work supported by the UK Science and Technology Facilities Council, the US Department of Energy, Office of Nuclear Physics, under Contract No. DE-AC02-06CH11357,  the National Science Foundation Grant No. PHY-08022648 (JINA) and the DFG Cluster of Excellence ``Origin and Structure of the Universe."


\end{document}